# Measuring the randomness of micro and nanostructure spatial distributions: Effects of Scanning Electron Microscope image processing and analysis

A. Mavrogonatos[1], E-M. Papia[1], P. Dimitrakellis[1] and V. Constantoudis[1,2,*]

[1]*Institute of Nanoscience and Nanotechnology, NCSR Demokritos, Agia Paraskevi, 15341 Greece*

[2]*Nanometrisis p.c., Agia Paraskevi, 15341 Greece*

*[*]v.constantoudis@inn.demokritos.gr*

**Abstract**: The quantitative characterization of the degree of randomness and aggregation of surface micro and nanostructures is critical to evaluate their effects on targeted functionalities. To this end, the methods of Point Pattern Analysis (PPA), largely used in ecology and medical imaging, seem to provide a powerful toolset. However, the application of these techniques requires the extraction of the point pattern of nanostructures from their microscope images. In this work, we address the issue of the impact that Scanning Electron Microscope (SEM) image processing may have on the fundamental metric of PPA, i.e. the Nearest Neighbour Index (NNI). Using as examples two typical SEM images of polymer micro- and nanostructures taken from secondary and backscattered electrons, we report the effects of the a) noise filtering and b) binarization threshold on the value of NNI as well as the impact of the image finite size effects. Based on these results, we draw conclusions for the safe choice of SEM settings to provide accurate measurement of nanostructure randomness through NNI estimation.

## 1. Introduction

The recent revolution of nanotechnology has been mainly based on two pillars. The first is the proliferation of techniques enabling the nanostructuring of material surfaces, which include both top-down and bottom-up approaches. The second pillar is the outstanding advancement of diagnostic techniques sensitive to nanoscale structures and especially of microscopy methods providing high resolution images of nanoworld. Nanostructuring endows materials with new properties and functions, which may catalyze several novel applications of nanotechnology, while microscopy and other diagnostic techniques boost deeper understanding and enable control of nanostructuring processes towards their scale-up in manufacturing production lines [1].

Usually, the outcome of a nanostructuring process can be either a continuous surface texture which is usually called surface roughness or a spatial distribution of discrete nanofeatures with well-defined borders (edges) such as nanodots, nanoholes, nanoparticles, nanowires or nanoflakes. In several cases where nanofeatures are strongly assembled, the surface can be considered either continuous or discrete depending on the angle of view we approach it. When the surface exhibits a continuous morphology, the theoretical tools of roughness characterization can be employed to provide quantitative metrics of surface morphologies consisting of both vertical (height distribution, rms, skewness, etc) and spatial (Fourier spectra, correlation functions, fractal/multifractal analysis) aspects [2-4]. On the other side of discrete morphologies, the couple of metrological demands transforms to the measurement of feature size distribution and the characterization of spatial organization of the positions of nanofeatures. In most cases, the focus is on the feature size distribution where several techniques and analysis methods have been employed [4-6]. However, much less work has been devoted to the quantitative characterization of the spatial arrangement of nanostructure positions on surfaces, despite the critical effects it may have on surface and film functionality such as mechanical stability, catalysis, antibacterial effects and optical transparency and reflectance to name just a few [7-9].

The quantitative characterization of spatial patterns has also attracted a lot of interest in other areas of science such as ecology, medical imaging and materials science [10-12]. The mathematical methodology enabling the spatial pattern characterization is usually called Point Pattern Analysis (PPA) and is considered a part of the stochastic geometry [13-15]. PPA outputs parameters and metrics to describe different aspects of spatial organization. Usually, the delivered results can be separated in first and second order metrics, depending on whether it is limited to the nearest neighbor distance or extend to spatial correlations besides this limit. The Nearest Neighbor Index (NNI) and Ripley's K-function are representative examples of metrics of 1st and 2nd order statistics respectively.

The estimation of these parameters requires to have first defined the point pattern. In the case of nanostructures, what we usually have available are top-down SEM pictures which display in grey scale the surface features. Commonly, two types of SEM images can be obtained, depending on whether the secondary and backscattered electrons are collected. The first type of secondary electron images are sensitive to surface morphology, whereas the backscattered signal depicts differences in the chemical synthesis of imaged surface. In both cases, an image processing step should be preceded so that we extract a point pattern from the analyzed SEM image. The question that naturally arises is whether and at which degree the intervening image processing steps are affecting the output PPA first and second order metrics.

In this paper, we address this issue focusing on the effects of two fundamental image processing steps, namely noise filtering and binary conversion, on NNI. Furthermore, we consider the impact of the limited range of SEM images on NNI and we compare the weighted edge correction and guard area techniques, two methods widely used in literature, paying special emphasis on run time issues in view of analysis of images with numerous nanostructures or other objects.

To meet the above targets, the paper is structured as follows. In the next section 2, we summarize the basic concepts of PPA with emphasis on their application in nanostructure analysis and present the basic steps of our methodology. Section 3 describes the images which will be analyzed along with some information about the nanostructures they exhibit. Then Section 4 presents the results of our analysis divided in subsections for the effects of noise filtering, of the threshold for BW conversion and of the comparison of edge correction algorithms. The paper closes with a summary of the derived conclusions in Section 5.

## 2. Methodology
### 2.1 Point pattern analysis – A short introduction

A point pattern is defined as a collection of N points ($p_1,p_2,…,p_N$) distributed in some region A. The region A may have any dimension and shape while each point is determined by some set of coordinates accordingly. In the case of nanotechnology applications, the points can be the centers of nanostructures (e.g. nanodots, nanoparticles, nanowires,etc) and the dimension of A may be 2 or 3, depending on whether we study nanostructure distributions on a surface or in the bulk of a material. In PPA literature, one can also find the term point process which albeit is not identical to the point pattern. A point process is usually a theoretic stochastic model generating point patterns, whereas the latter are actually realizations of point processes.

Furthermore, we can consider "marked" point patterns where an additional characteristic (mark) is anchored at each point, such as the diameter of nanoparticles or the length and orientation of nanowires in nanostructure point patterns.

The main question in PPA is to find ways to quantify the departure of a PP from the Complete Spatial Randomness (CSR) where the points of PP are randomly placed in area A with no correlations and interactions between them. Although there are several ways to realize the deviation from randomness, a useful classification is between patterns with more even dispersion of points closer to periodic lattices and patterns with strong aggregation tendencies. The first metric to provide such a quantification was proposed several decades ago in the seminal paper by Clarks and Evans[16] and it was based on the concept of nearest neighbor distances $d_i$, i=1,..,N , i.e. the distance of the i-th point from its nearest neighbor in a pattern with N points totally. The proposed metric is usually called Nearest Neighbor Index (NNI) and equals the ratio of the average nearest neighbor distance $d_A = <d_i>$ over the expected one $d_E$ when the hypothesis of CSR holds:

$$NNI = \frac{d_A}{d_E} = \frac{<d_i>}{0.5\sqrt{A/N}}$$

since it can be shown that in CSR $d_E = 0.5\sqrt{A/N}$ based on the fact that the points are distributed according to the Poisson distribution on area A [16].

NNI is a measure of the deviation of the observed point pattern from the CSR hypothesis and can easily quantify the degree of clustering in the spatial distribution of nanostructures. By definition, a point pattern distributed randomly shows *NNI*=1.0, while regularity (even spatial distributions) can be assumed if *NNI*> 1.0. When NNI< 1.0 then clustering effects in nanostructure distributions can be inferred. The maximum value of the index is reached for a strict hexagonal pattern where *NNI*= 2.15 [17, 18]. NNI is used as a preliminary step in the PPA and provides information about the point pattern in a fine spatial scale, since it is limited to the statistics of nearest neighbour distances scale.

Furthermore, as a distance-based metric, NNI suffers from the bias induced by the finiteness of the measurement image area. In the literature related to PPA, these effects are named edge effects, where the term edges means the borders of image area [13]. The origin of the bias is that the points near the area edges may have a closer neighbour outside the imaged region than the couple point found inside area. This leads to an overestimation (positive bias) of NNI if no correction algorithms are applied. One can find several such algorithms in the literature, ranging from periodic or reflection-based extensions of patterns to the more widely used border (or minus-sampling or guard) and nearest-neighbour (or weighted) methods [13]. In the border method, a subset of image area is considered with sides moved inwards by the largest nearest neighbour distance of points so that to get unbiased calculation of NNI though with poorer statistics. The nearest neighbour method is more sophisticated and considers in NNI calculation all points with nearest neighbour distance smaller than the lowest distance of the point from image borders. To compensate for the bias induced by the exclusion of points with relatively large nearest neighbour distance, the averages in NNI estimation are weighted according to their nearest neighbour so that points with large NND to have higher contribution. Due to its extensive use, the nearest-neighbour method will be called here the standard method for edge correction.

Besides the first-order statistical test of NNI, PPA has elaborated second-order statistical tools to quantify the spatial long-range correlations of point positions in pattern. The most popular is the cumulative Ripley's K-function which is defined as the average number of points within a predefined radius of any point, normalized for the point density over the whole image area. Along with K-function, its linear transform L(r) and its noncumulative (rescaled) derivative neighbourhood density function (analogous to the pair-correlation function) are also applied in many cases to disclose more detailed spatial information about the long range correlations and interactions among points.

In this paper, we are limited to the NNI alone due to its primary role in PPA and its easily implemented and fast calculation. The focus is on the investigation of the effects that the microscope image processing methods needed to extract the pattern of micro and nanostructure positions have on the value of NNI. Additionally, since SEM depicts a finite area of the whole micro and nanostructured surface, we study the impact of the more popular edge correction algorithms (border and nearest-neighbour) on the NNI estimation. In the same spirit, we are planning to extend our work to the examination of more detailed spatial metrics such as second order functions and especially the Ripley's K-function.

Our ultimate aim is to provide a sound basis to the application of the full toolset of PPA methods in the quantitative characterization of nanostructure dispersions in 2D (surface) and 3D (bulk) environments. We believe that the successful application of these methods will have a significant contribution to the quantitative characterization of nanostructured surfaces disclosing hidden aspects of their impact on nano-enabled functionalities.

## 2.2 Flowchart of our methodology

Fig.1 shows the flowchart of the methodology we followed to study the effects of image processing steps on NNI. We start with the grey-scale image as acquired by the SEM measurement and then we implement two possible scenarios for the processing of the image to extract the nanostructure point pattern. In the first, we first convert the grey-scale image to binary (black-white) by applying a properly chosen threshold. Then we reduce the noise in the binary image by applying a noise filter which can be linear (Gaussian) or nonlinear (median). In both cases, the extension of noise smoothing filter is determined by a parameter which can be the Gaussian filter width or the window of median filter.

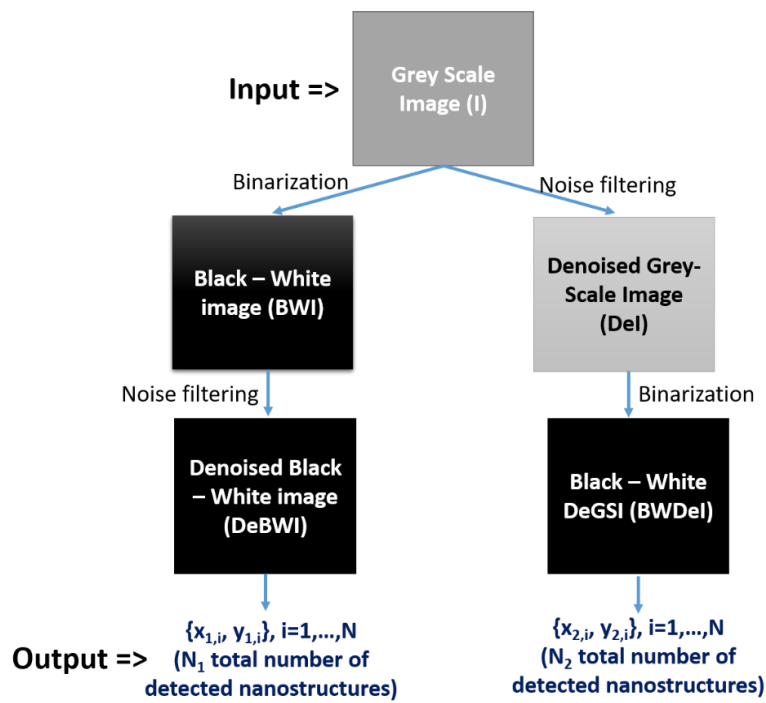

**Figure 1.** Flowchart of our methodology with the two scenarios for the processing of the SEM image to obtain the point pattern of nanostructure surface dispersions.

The inverse series of processes is applied in the second scenario, where the first step is the noise smoothing and the second the image binarization and conversion to a black-white image. In both scenarios, the involved parameters are similar (noise filter width or window and binarization threshold) as well as the output set of nanostructure centers which plays the role of point pattern used in the estimation of NNI.

Therefore, the aim of our study is specified to quantify the effects of a) noise smoothing filter width or window and b) threshold value on the calculated NNI. Furthermore, in the estimation of NNI we consider biased approaches with no edge correction along with edge-correction algorithms, based either on weighted or guard methods.

## 3. Description of data

Recently, we have shown that atmospheric pressure plasma etching of the photopolymer resist AZ5214E coated on Silicon leads to the formation of well-defined isolated residuals on the underlying substrate with micro and nanoscale [19]. Herein, we analyze two top-down images acquired using a FEI Quanta Inspect Scanning Electron Microscope (SEM) which depict the plasma-assembled structures at different samples. The first has been obtained from a thin photoresist film ( ~ 650 nm) using the Secondary Electron mode (SE) (Fig. 2a) and the second from a thicker film ( > 1 μm) using the Back-scattered Electrons (BSE) mode (Fig.2b). The Silicon substrate in both cases was much thicker reaching 380μm.

The atmospheric pressure plasma etching was conducted in a parallel-plate, showerhead-type dielectric barrier discharge (DBD) source operating at radio- frequency (RF) 13.56 MHz [19-23]. The experiments were performed in open-air environment, while the process gases, namely helium and oxygen, were injected through the showerhead electrode to the plasma region at a total flow rate of 5 L/min. The oxygen molar fraction was fixed at 0.6 %, the RF power was 120 W and the electrode gap 2 mm. The samples were placed on the surface of the water-cooled grounded electrode. The process was also pulsed (5 s plasma-on / 15 s plasma-off) in order to prevent over-heating.

We limit the presentation of the results of our study (see Section 4) to these images because they are typical representatives of the family of SEM images showing nanostructures on surfaces either in secondary or back-scattered electron mode of SEM operation. More images have been analysed indicating similar trends and justifying the drawn conclusions.

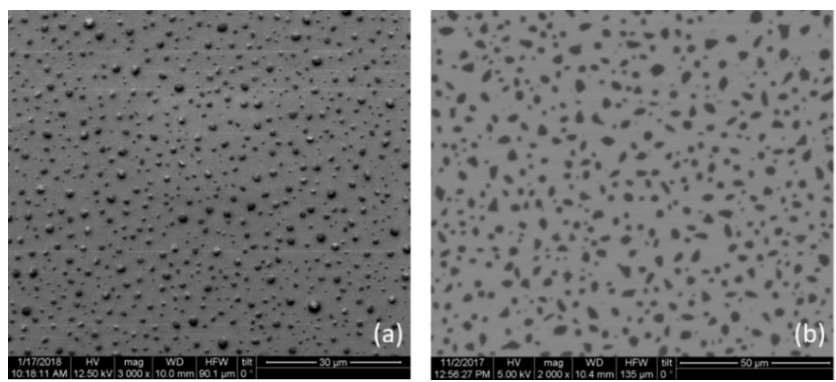

**Figure 2**: Top-down SEM images of polymer AZ micro and nanoresiduals on silicon substrates acquired using secondary (a) and backscattered (b) electron signals.

## 4. Results

As explained in Methodology section, the results of our study concern the impact of image processing parameters (filter width and binarization threshold) on NNI as well as the effects (if any) that the order of the two fundamental image processing steps (noise filtering and thresholding) may have. We are also interested in the role of the finite size of image area on NNI and compare two edge correction algorithms with the straightforward but biased results. To make the presentation of the findings more clear, we organize this section in four subsections each one dedicated to a particular dependence of NNI (noise filter, threshold, order of steps and edge correction).

### 4.1 Impact of noise filter

One of the main obstacles to extract useful and reliable information from an image is the presence of noise which can overwhelm the strength of the real signal and lead to confused results. This is why the application of smoothing filters to reduce the harmful noise effects is of first priority in image processing techniques. In the case of SEM images, the noise has two main sources. The first is the Poisson distribution of the intensity of incident electron beam which results in similar distribution in the obtained output signal (SE or BSE) and the second is Gaussian-like since it comes from the electronic noise of SEM circuits which transform the received signal into image pixel intensity.

In literature, one can find both linear and nonlinear filters to reduce noise effects. When linear filters are used, each image pixel is replaced by a weighted average over its neighbor pixels. The most usual choice for the weighted function is the 2D Gaussian function which is defined by its width sigma.

Figure 3a shows the impact of the width of a Gaussian noise filter on the estimated NNI for both the SE (red line) and BSE (blue line) images. The standard approach to remedy the finite image area effects (edge correction) has been applied in both calculations, while the binarization is kept fixed equal to 0.5. The processing steps are also ordered with first the noise smoothing and then the conversion to black and white image.

The NNI of BSE image remains almost fixed versus Gaussian width exhibiting a very slight increase from1 .42 to 1.44 when width goes from 0.5 to 4. The magnitude of NNI means that the spatial pattern of imaged micro and nanostructures declines clearly from CSR in favor of periodicity and homogeneity. A similar behavior is indicated by the NNI of SE image with the difference that the saturation is reached when noise smoothing exceeds a critical value of width about 1.2. The saturated value is marginally smaller than 1.4 which means than the imaged micro and nanostructures have a clear tendency towards periodicity though it is slightly weaker than that found in BSE image.

The above-mentioned difference between BSE and SE images at small filter widths is due to the relative strength of their signals. The signal of SE is much weaker than that of BSE and therefore it is more vulnerable to noise effects. Therefore SE images are suffered more from the presence of noise and require first to apply a smoothing filter and then proceed to binarization and point pattern extraction.

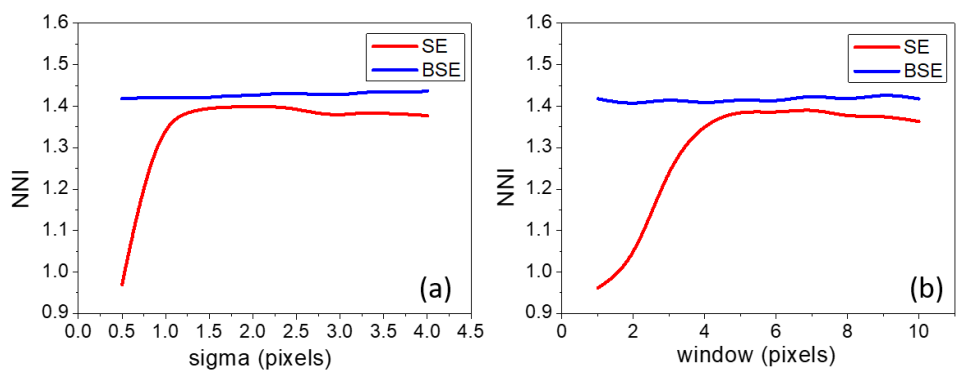

**Figure 3.** Gaussian (a) and median (b) noise filter effect on NNI for both SE (red line) and BSE (blue line) images. Notice that we have applied the standard edge correction method in both cases.

Besides linear filter, we have also applied nonlinear filters and particularly the median filter to reduce the image noise. The great benefit of nonlinear filters is that they are edge-preserving, i.e. in most cases they do not degrade the sharpness of edges defining the borders of imaged structures. Therefore, nonlinear filters seem to be a more appropriate choice when we want to measure the size of micro and nanostructures on images and extract the coordinates of their centres to obtain a marked point pattern. Although this is not the aim of this paper, we include the effects of the window of a median filter on NNI for the sake of completeness. Figure 3b shows the results of our analysis for both images and windows ranging from 1 to 10. Similarly to the effects of Gaussian filter, the NNI of BSE image is almost unchanged with window size whereas in the case of SE image the NNI saturates when the window is larger than 4. The interpretation of this behaviour is similar to that articulated in linear filters and is based on the different amount of noise in BSE and SE images.

The final output of this subsection is that the linear and nonlinear noise smoothing filters have different impact on the estimation of NNI in SE and BSE image due to their different noise content. In SE images, the reliable estimation of NNI requires first the noise smoothing step which can be done through linear or nonlinear filters with similar results. Furthermore, the presence of noisy pixels randomizes the detected point pattern and this is the reason of the decrease of NNI at small noise smoothing parameters.

### 4.2 Impact of binarization threshold

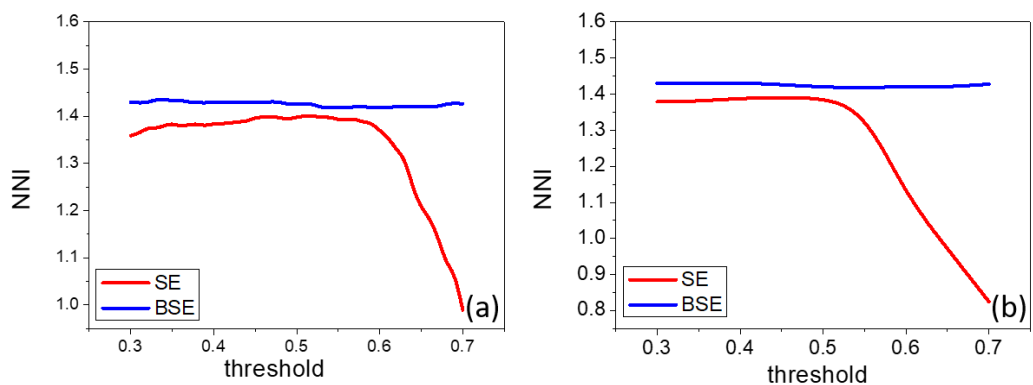

**Figure 4.** Effects of binarization threshold on NNI for fixed Gaussian (a) and median (b) noise filtering. The behavior of both SE (red line) and BSE (blue line) images are shown while the standard edge correction method has been applied.

The conversion of the SEM grey scale images to black and white is a prerequisite step towards the micro and nanostructure identification on the image, the determination of their area and centers and finally the extraction of the desired point pattern. The binarization is usually made through the choice of a threshold replacing all pixels with intensity greater than threshold with 255 (white) and all other pixels with the value 0 (black). The threshold definition is a quite complicated issue especially when the histogram of image is unimodal with no clear effects of micro and nanostructures on it (the case of SE image). In order to quantify the threshold effects on NNI, we present the diagrams of Figure 4 which illustrate the NNI versus threshold when a Gaussian (Fig.4a) or a median (Fig.4b) is applied to reduce noise effects. The NNI of BSE image in both cases remains almost fixed for all thresholds, in harmony with the results of the previous subsection and the more well-defined regions of micro and nanostructures. Contrary to that, SE image is characterized by an NNI which changes with threshold falling down when threshold becomes greater than ~0.6 for Gaussian filter and ~0.5 for median filter. The reason of this reduction is the vanishing of the structure areas with medium intensity values when threshold increases towards higher values. Since the positions of these structures are expected to be random, their removal results in more random patterns of imaged structures and therefore in NNI values with a bias towards 1 which the expected value for CSR.

### 4.3 Impact of the ordering of noise filtering and binarization

As explained in the Methodology Section 2, the reliable extraction of a point pattern from a SEM image requires the application of two fundamental image processing techniques: the noise filtering and the conversion of the image to a binary format. In this subsection, we pose the question if the final results of our PPA, i.e. NNI value, is sensitive to the order of the application of these techniques. The motivation of this inquiry is the nonlinear character of the binarization and denoising process when the median filter is used.

Fig. 5 shows the similar diagrams to Fig. 3b and 4b but when we apply the reverse order of image processing processes, i.e. first the binarization and then the noise smoothing filter. We prefer to show the case of median filter to take into account its nonlinearity. One can easily notice the similarity of NNI dependencies of Fig.5a and Fig. 5b with those of Fig.3b and Fig.4b respectively. This result justifies the insensitivity of NNI calculation to the order that image denoising and binarization.

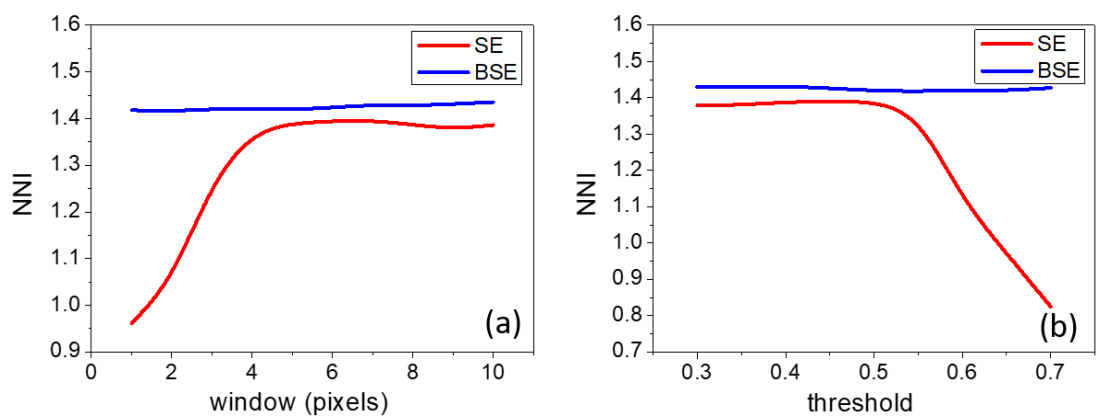

**Figure 5.** NNI vs. median noise filtering window (a) and binarization threshold (b) for both BSE (blue line) and SE (red line) images when we apply first binarization and then noise smoothing filter.

### 4.4 Edge correction

As we explained in the Introduction, the extraction of a point pattern from an SEM image and the concomitant estimation of NNI should consider the edge effects of the finite area of SEM image. This subsection is dedicated to explore the effectiveness of the border and nearest-neighbour methods to remedy edge effects in the SEM images studied in this paper. For the sake of comparison we also include in the diagrams the biased values of NNI calculated with no edge correction. We remind that the results of the previous sections have been obtained using the standard method for edge correction (nearest neighbour method).

Fig. 6 shows the NNI values for SE image (Fig. 6a) and BSE image (Fig. 6b) for threshold=0.5 and median filter with a window=6 to have sufficient removal of noise effects. As we expected, the calculation of NNI with no edge correction leads to a biased overestimated NNI in both images. The application of border method lowers slightly NNI estimation towards the NNI of the standard method which is clearly smaller than both biased and border predictions. Similar tendencies have been found when Gaussian filter or other thresholds are used in the extraction of point pattern. Although the bias in NNI value is smaller than 10% (due to the strong periodic contribution of SEM point patterns), the application of the standard nearest-neighbour edge correction algorithm is recommended since it provides an easily implemented method to get more accurate results.

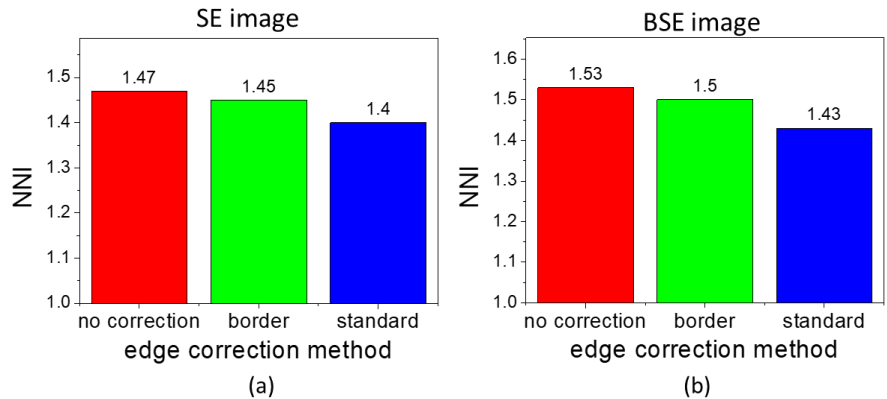

**Figure 6**. Bar diagram with the values of NNI calculated with no edge correction, border and standard correction methods for SE (a) and BSE (b) images.

## 5. Summary – Conclusions

Despite the significant role the distribution of micro and nanostructures on surfaces and bulk materials plays in their properties and functionalities, a concise methodology for the quantitative characterization of its spatial aspects is still missing. In this paper, we propose the application of PPA methods to fill this gap and accelerate the implementation of these structures in industrial environments. However, first we should adapt these methods to the specific characteristics of microscope images used for the extraction of the point patterns of micro and nanostructures.

To this end, we start with a thorough investigation of the impact of image processing and analysis steps of denoising and binarization on the value of the primary PPA metric of NNI. To get realistic results, we select two typical experimental SEM images of polymer micro and nanostructures acquired using backscattered and secondary electron output signal for testbed of the dependencies and methods. Moreover, we examine the effects that the order of the application of these image processing steps as well as the correction of the bias induced by the finite area of SEM images may have on NNI.

The findings of our paper can be enumerated as follows:

1. The effects of denoising window and binarization threshold on NNI depend on the amount of noise in the analyzed image. The estimation of NNI in BSE images, which normally have low noise level, is quite robust to the variations of these parameters. Contrary, in the noisier SE images, we should pay attention to select the proper noise smoothing filter width and binarization threshold to get a reliable calculation of NNI based only on the real micro and nanostructure positions. Otherwise, biased results are obtained which are poisoned by the presence of noise or the random elimination of small structures.

2. The order that the two steps of denoising and binarization are applied to get the desired point pattern seems to be irrelevant to the output NNI value.

3. The finite range of SEM images affects the calculated NNI value in both SE and BSE images. Therefore, it is strongly recommended the application of the standard edge correction algorithm before analysis to remedy these effects and provide more accurate NNI measurements. The border method, though it corrects the NNI towards the right direction, does not provide sufficient correction.

The next steps of our work will be devoted to the investigation of image processing and analysis steps on more advanced NNI metrics such as the Ripley K-function and the related L-function as well as the pair correlation function. We aim to generate synthesized SE images to get a full exploration of the effects of noise, spatial nanostructure size and SEM image area on PPA results.

After the establishment of the PPA approach in the characterization of micro and nanostructure dispersion on surfaces and materials, we will consider marked point patterns of micro and nanostructures taking into account critical non-spatial features such as their size or orientation. Finally, the enrichment of our nanometrology toolbox with PPA techniques should be related to the micro and nano-enabled properties and functionalities, revealing its potential to provide better control of the link between structure and functionalities in nanotechnology. The reinforcement of this link is critical to the scale-up micro and nanostructure fabrication in industrial manufacturing processes.